\documentclass[showpacs,preprintnumbers,amsmath,amssymb]{revtex4}
\usepackage{amsmath}
\usepackage{graphicx}
\usepackage{subfig}
\usepackage{hyperref}
\usepackage{amssymb}
\usepackage[english]{babel}
\usepackage{epsfig}
\usepackage{wasysym}

\begin{document}
\title{Spherical Accretion of Matter by Charged Black Holes on f(T) Gravity\\}
\author{Manuel E. Rodrigues$^{(a,b)}$}\email{esialg@gmail.com}
\author{Ednaldo L. B. Junior$^{(b,c)}$}\email{ednaldobarrosjr@gmail.com}
\affiliation{$^{(a)}$ Faculdade de Ci\^{e}ncias Exatas e Tecnologia, 
Universidade Federal do Par\'{a}\\
Campus Universit\'{a}rio de Abaetetuba, CEP 68440-000, Abaetetuba, Par\'{a}, 
Brazil} 

\affiliation{$^{(b)}$ Faculdade de F\'{\i}sica, PPGF, Universidade Federal do 
 Par\'{a}, 66075-110, Bel\'{e}m, Par\'{a}, Brazil}

\affiliation{$^{(c)}$ Faculdade de Engenharia da Computa\c{c}\~{a}o, 
Universidade Federal do Par\'{a}, Campus Universit\'{a}rio de Tucuru\'{\i}, CEP: 
68464-000, Tucuru\'{\i}, Par\'{a}, Brazil}

\begin{abstract}
We studied the spherical accretion of matter by charged black holes on $f(T)$ Gravity. Considering the accretion model of a isentropic perfect fluid we obtain the general form of the Hamiltonian and the dynamic system for the fluid. We have analysed the movements of an isothermal fluid model with $p=\omega e$ and where $p$  is the pressure and $e$ the total energy density. The analysis of the cases shows the possibility of spherical accretion of fluid by black holes, revealing new phenomena as cyclical movement inside the event horizon.
\end{abstract}

\maketitle


\section{Introduction}
\label{sec1}
Gravitation and Cosmology modern are based on General Relativity (GR), which in turn is based on the riemannian Differential Geometry. The structure of the so-called space-time, a four-dimensional differentiable manifold, is completely characterized by a metric, because the connection,  Riemann and Ricci tensors and scalar curvature depend only of the metric components, its inverse and their derivatives \cite{wald}. It is a fact that this structure can only describe the current evolution of our universe, being consistent with the observational data, introducing a fluid model where the pressure must necessarily be negative today, model known by $\Lambda$CDM \cite{frieman}. 
\par 
An alternative description for the accelerated evolution of our universe is one in which modifies the Einstein equation, for a generalized such that at some threshold parameters theory falls in GR. The simplest way to do this is to directly modify the Einstein-Hilbert action to another which has the limit of GR, such as of the $f(R)$ Gravity \cite{fR}, which can be made the limit for small curvatures and get $f(R)\sim a_0+a_1R+O(R^2)$, for example. It is common generalize GR by modifying the Einstein-Hilbert action, using a function of a scalar theory, as the theories $f(R,\mathcal{T})$ Gravity \cite{fRT}, $f(G)$ \cite{fG} and $f(R,G)$ \cite{fRG} Gravities, for example, where $\mathcal{T}$ is the trace of the energy-momentum tensor and  $G$ the Gauss-Bonnet term.
\par 
An alternative description, but equivalent, of gravity can be done now using the torsion of space-time, rather than curvature. Considering now the identically zero curvature, we can assign the gravitational interaction effects to torsion. The theory that represents such an alternative is commonly called Teleparallel Theory (TT) \cite{TT}, where are now tetrads that make the role of dynamic fields, which completely determining the geometric objects such as connection, torsion, contorsion and Torsion scalar. As this theory is equivalent to GR, the equations of motion can only describe the accelerated evolution of the current phase of our universe through again the introduction of dark energy. Then we can think analogously to generalize this theory to one that contains terms of order higher torsion scalar, as $f(T)$ Gravity \cite{fT}. We also have the same possibilities for generalization for $f(T,\mathcal{T})$ \cite{fTT} and $f(T,T_G)$ \cite{fTG} Gravities, or more generally as in \cite{gonzalez}. In this paper, we will restrict ourselves to $f(T)$ Gravity.
\par 
In the case of the study of Cosmology by $f(T)$ Gravity, we have the most varied results. The study by local phenomena such as black holes solutions, we still have much work ahead of us. The first hole black solution theory is charged solutions on $3D$ \cite{gonzalez2} and $D$ dimensions \cite{capozziello1}. Later appears more charged solutions in $4D$, re-obtaining thus solutions of Reisnner-Nordstrom, Reisnner-Nordstrom-AdS (dS), Schwarzshild-AdS (dS) e Schwarzshild \cite{rodrigues1}. Another paper discusses the formulation of the Kerr solution for $f(T)$ Gravity \cite{bejarano}. The solutions of charged black holes, resulting in non-linear electrodynamics are obtained in \cite{junior}, which we will study here in this work. We also have regular black holes solutions in \cite{junior2}. Even with few solutions found in this theory, we have very few studies of local phenomena such as bending of light \cite{ruggiero} and solar tests \cite{farrugia}.
\par 
The study of accretion of matter by a black hole was first performed by Bondi \cite{bondi}, but in a newtonian form. Generalization to account for relativistic effects was formulated by Michel on 1971 \cite{michel}, so is called accretion type-Michel for this approach. In this paper we will use the accretion type-Michel. The description of spherical accretion made by Michel is the phenomenon critically, having a certain radius value where the system becomes critical. Considering effects of curvature, we have several results of spherical accretion by black holes \cite{accretion1}, and some are for modified gravity \cite{accretion2}. The only pioneer work in considering the zero curvature and the torsion effects on $f(T)$ Gravity was recently published \cite{accretionfT}. We will follow the same methodology of this work but generalizing the accretion analysis for a general spherically symmetric metric.
\par 
So our motivation is to treat the spherical accretion of matter by charged black hole of the $f(T)$ theory. In Section \ref{sec2} we discussed the spherical accretion of matter by a black hole, giving the expression of the Hamiltonian of the fluid and dynamic system. In Section \ref{sec2} we restrict our analysis to the model of isothermal isentropic fluids, particularizing for ultra-stiff cases, ultra-relativistic, radiation and sub-relativistic fluids. In Section \ref{sec3} we present our final considerations.          


\section{Spherical Accretion}
\label{sec2}
The spherical accretion study of matter by black holes is usually based on the movement of fluid in the neighboring region to the event horizon. This can be addressed as follows. Let's take two equations that completely characterizes the accretion (ejection) spherical of a perfect isotropic fluid; one is the equation of conservation of source material and the other is the equation that describes the conservation of energy. Let's start by the equation similar to that of continuity in fluid mechanics, which is described by \cite{zanotti} 
\begin{eqnarray}
\nabla_{\mu}J^{\mu}\equiv 0\label{j}\,,
\end{eqnarray}
where $J^{\mu}=nu^{\mu}$, $n=\rho$ is the number of the baryonic density and  $u^{\mu}=dx^{\mu}/d\tau$ is the four-speed fluid.

Analogously to the theorem of Noether of fluid mechanics, Bernoulli's theorem in hydrodynamics relativistic have to spherical symmetry, a certain amount that is conserved in the co-moving frame to fluid flow. This law is represented by the equation \cite{zanotti}
\begin{eqnarray}
u^{\nu}\nabla_{\nu}\left[hu_{\mu}\zeta^{\mu}\right]\equiv 0\label{Bernoulli}\,,
\end{eqnarray}   
where $h$ is the enthalpy of the fluid and  $\zeta^{\mu}$ a Killing vector of the temporal symmetry of space-time. 
\par 
To simplify our study and make the equations having analytical solutions, we will restrict ourselves to the case where the space-time has spherical symmetry, which implies in the following general form of the line element
\begin{eqnarray}
dS^2=A(r)dt^2-B(r)dr^2-C(r)\left(d\theta^2+\sin^2\theta d\phi^2\right)\label{ele}\,.
\end{eqnarray}
For the line element \eqref{ele} we have the following metrics determinant  $g=-ABC^2$. 

The fluid movement follows the same symmetries of space-time, ie spherical symmetry. So we can then take a fluid that moves in temporal and radial directions, and its four-speed given by $[u^{\mu}]=[u^0,u,0,0]$. Thus, because of the spherical symmetry, the continuity equation \eqref{j}, integrated, gives us
\begin{eqnarray}
C\sqrt{AB}nu=c_1\label{c1}\,.
\end{eqnarray}
This is one of the equations principles that govern the movement of the fluid. 

We can then normalize the four-speed for $g_{\mu\nu}u^{\mu}u^{\nu}=1$, which gives us the following identities
\begin{eqnarray}
u^{0}=\sqrt{A^{-1}\left(1+Bu^2\right)}\,,\,u_{0}=g_{00}u^{0}=\sqrt{A\left(1+Bu^2\right)}\,.\label{u0-0}
\end{eqnarray}

Because of spherical symmetry, it is common we take the following vector of Killing $[\zeta^{\mu}]=[1,0,0,0]$ to integrate \eqref{Bernoulli}, resulting in
\begin{eqnarray}
h\sqrt{A\left(1+Bu^2\right)}=c_2\label{c2}\,.
\end{eqnarray}

We may relate the three-velocity of the fluid with the radial and temporal components of the four-velocity through the line element to the equatorial plane $\theta=\pi/2$
\begin{eqnarray}
dS^2(\theta=\pi/2)=\left(\sqrt{A}dt\right)^2-\left(\sqrt{B}dr\right)^2\label{ele3}\,.
\end{eqnarray}
Now we define the three-velocity of the fluid by
\begin{eqnarray}
v=\frac{\sqrt{B}dr}{\sqrt{A}dt}\,,
\end{eqnarray}
that replacing $u=dr/d\tau$ and $u^{0}=dt/d\tau$, we have
\begin{eqnarray}
v^2=\frac{B}{A}\left(\frac{u}{u^{0}}\right)^2\label{eq1}\,.
\end{eqnarray}
Using \eqref{u0-0} and isolating $u^2$, we have
\begin{eqnarray}
u^2=\frac{v^2}{B(1-v^2)}\,,\,u_{0}^2=\frac{A}{1-v^2}\label{u0}\,.
\end{eqnarray}

Now through \eqref{u0} we can rewrite our equation \eqref{c1} as
\begin{eqnarray}
\frac{A\left(Cnv\right)^2}{1-v^2}=c_1^2\label{c1-1}\,.
\end{eqnarray}

Bernoulli's theorem of relativistic hydrodynamics says that every symmetry of space-time we have a conserved quantity associated with this symmetry. For the temporal translational symmetry, the conserved quantity is the square of the integration constant $c_2$ in \eqref{c2}, and should be proportionate to the Hamiltonian of the fluid. using again \eqref{u0}, we define the Hamiltonian fluid as
\begin{eqnarray}
\mathcal{H}(r,v)=\frac{h^2(r,v)A(r)}{1-v^2}\label{H0}\,.
\end{eqnarray} 

The intensive and extensive thermodynamic quantities of the fluid are related by equations
\begin{eqnarray}
dp=n\left(dh-Tds\right)\,,\,de=hdn+nTds\,,
\end{eqnarray}
where $T$ and $s$ are specific entropy and the temperature of the fluid, and  $e$ is the total energy density of the fluid. In general, the study of fluid motion becomes very complicated, and it is due to this reason that we must assume two simplifications for perfect fluid our case. The first is considered that a fluid does not heat exchange with the outside, then the fluid is classified as adiabatic fluid, and  $u^{\mu}\nabla_{\mu}s\equiv 0$. The second is when the fluid keeping its entropy constant during movement, then the fluid thus characterized by isentropic where $\nabla_{\mu}s\equiv 0$ and therefore $s=s_0\in\Re^{+}$. 

Considering that our case study is a fluid adiabatic and isentropic, this implies $ds\equiv 0$ and then 
\begin{eqnarray}
dp=ndh\,,\,de=hdn\label{termo}\,.
\end{eqnarray}  
We can divide $dp$ by $de$ to define the speed of sound as
\begin{eqnarray}
a^2=\frac{dp}{de}=\frac{d\ln h}{d\ln n}\label{a}\,.
\end{eqnarray}

Now we will establish a dynamic system through Hamiltonian \eqref{H0}
\begin{eqnarray}
\frac{dr}{dt}=\frac{\partial \mathcal{H}}{\partial v}\,,\,\frac{dv}{dt}=-\frac{\partial \mathcal{H}}{\partial r}\label{dinamic0}
\end{eqnarray} 

Now do the same steps contained in \cite{accretionfT}, remembering that our metric is generalized to \eqref{ele}, what gives us the following dynamic system
\begin{eqnarray}
&&\dot{r}=\frac{2Ah^2}{v(1-v^2)^2}(v^2-a^2)\,,\label{dotr}\\
&&\dot{v}=-\frac{h^2}{1-v^2}\left[\frac{dA}{dr}-2a^2A\frac{d\ln(\sqrt{A}C)}{dr}\right]\,.\label{dotv}
\end{eqnarray}

We have here some possibilities of critical points for this dynamic system, but we will focus only on single physical possibility, where
\begin{eqnarray}
v_c=a\,,\,A'=2a^2A(\ln \sqrt{A}C)'\,,\label{critical}
\end{eqnarray}
which defines a critical point for the system \eqref{dotr}-\eqref{dotv}.

We will now specify a famous model for the perfect fluid for the next section.

\section{Isothermal Fluids}

An important model that is well in agreement with the reality is when the fluid keeps a constant temperature in the thermodynamic process and thus the speed of sound is a constant, $a^2=\omega=dp/de$. This is another very important simplification to analytically solve the equations of motion of the fluid. Thus, integrating, we have what the pressure is proportional to total energy density 
\begin{eqnarray}
p=\omega e\label{pf}\,.
\end{eqnarray}

We define the enthalpy of the fluid by \cite{zanotti,accretionfT}
\begin{eqnarray}
h=\frac{e+p}{n}\label{h0}\,,
\end{eqnarray}
Here we can find an ordinary differential equation for the total density of the fluid, for this, we take \eqref{termo}, \eqref{pf} and \eqref{h0}, we integrate with respect to $n$, which provides us 
\begin{eqnarray}
e(n)=\frac{e_c}{n_c^{\omega+1}}n^{1+\omega}\label{ef}\,.
\end{eqnarray}
The enthalpy \eqref{h0} is then given by
\begin{eqnarray}
h=\frac{(1+\omega)e_c}{n_c^{\omega+1}}n^{\omega}\label{hf}\,.
\end{eqnarray}

We want to work with the thermodynamic variables $(r,v)$, then we must put the other thermodynamic quantities on the basis of these two variables. We then replace the dependence on $n$ for $n(r,v)$, for this, we will use \eqref{c1}
\begin{eqnarray}
n=\frac{c_1}{vC}\sqrt{\frac{1-v^2}{A}}\label{n}\,.
\end{eqnarray}
We now take \eqref{hf} and \eqref{H0}, considering \eqref{n}, to rewrite as
\begin{eqnarray}
\mathcal{H}=\mathcal{H}_0\frac{1}{\left(vC\right)^{2\omega}}\left[\frac{A}{1-v^2}\right]^{1-\omega}\,,\,\mathcal{H}_0=\left[\frac{(1+\omega)e_cc_1^{\omega}}{n_c^{1+\omega}}\right]^2
\end{eqnarray}
We can then define $\mathcal{H}$ by a transformation in the time coordinate (Killing vector of this symmetrical), or $t\rightarrow\mathcal{H}_0t$, resulting in $\mathcal{H}\rightarrow \mathcal{H}/\mathcal{H}_0$, then we redefine the Hamiltonian for
\begin{eqnarray}
\mathcal{H}[r,v]=\frac{1}{\left[vC(r)\right]^{2\omega}}\left[\frac{A(r)}{1-v^2}\right]^{1-\omega}\,.\label{Hf}
\end{eqnarray}

In the next subsection we will specify the cases between the relationship of pressure and total energy density of the fluid.

\subsection{Ultra-stiff fluid}\label{subsec1}
The model in which $\omega=1$ is called Ultra-stiff fluid. Thus, the Hamiltonian \eqref{Hf}, for this case becomes
\begin{eqnarray}
\mathcal{H}=\frac{1}{r^4v^2}\,,\,v_c=\sqrt{\omega}\label{H2}
\end{eqnarray}

The behaviour of the dynamic system \eqref{dotr}-\eqref{dotv} is the same for each specified solution. We will take here the following solutions of charged black hole for $f(T)$ Gravity \cite{junior}
\begin{eqnarray}
&&dS_1^2=\left(1-\frac{2r_0}{r^2}\right)dt^2-\left(1-\frac{2r_0}{r^2}\right)^{-2}dr^2-r^2\left(d\theta^2+\sin^2\theta d\phi^2\right)\,,\,^{(1)}F^{10}=-\frac{4r_0^2}{qr^3}\sqrt{1-\frac{2r_0}{r^2}}\,,\label{sol1}\\
&&dS_2^2=\frac{1}{r^2}\sqrt{r^4-4e^{r_0}}dt^2-\frac{r^8}{(r^4-4e^{r_0})^2}dr^2-r^2\left(d\theta^2+\sin^2\theta d\phi^2\right)\,,\,^{(2)}F^{10}=-\frac{16e^{2r_0}}{qr^8}\left(r^4-4e^{r_0}\right)^{3/4}\,,\label{sol2}\\
&&dS_3^2=\frac{1}{r^2}\sqrt[3]{r^6-2r_0}dt^2-\frac{r^{12}}{(r^6-2r_0)^2}dr^2-r^2\left(d\theta^2+\sin^2\theta d\phi^2\right)\,,\,^{(3)}F^{10}=-\frac{4r_0^2}{qr^{12}}\left(r^6-2r_0\right)^{5/6}\,,\label{sol3}\\
&&dS_4^2=\left(e^{r_0/2}r-1\right)^{-2}\left(dt^2-dr^2\right)-r^2\left(d\theta^2+\sin^2\theta d\phi^2\right)\,,\label{sol4-1}\\
&&^{(4)}F^{10}=-\frac{r(e^{r_0/2}-2)^2}{q(e^{r_0/2}r-2)^3}\left(2-4e^{r_0/2}r+e^{r_0}r^2\right)\left(e^{r_0/2}r-1\right)^2e^{r_0/2}\,,\label{sol4-2}
\end{eqnarray}
where $^{(i)}F^{10}$ with $i=1,...,4$ are the Maxwell tensor components. The event horizons are determined by $B^{-1}(r)=0$, that through \eqref{sol1}-\eqref{sol4-1} are given by 
\begin{eqnarray}
r_{h(1)}=\sqrt{2r_0}\,,\,r_{h(2)}=\sqrt{2}e^{r_0/4}\,,\,r_{h(3)}=\left(2r_0\right)^{1/6}\,,\,r_{h(4)}=e^{-r_0/2}\label{horizon}\,.
\end{eqnarray}

These are four solutions of charged black holes arising from $f(T)$ theory, with non-linear electrodynamics source obtained in \cite{junior}. In the case of ultra-stiff fluid all solutions behave likewise, this results already indicated in \cite{accretionfT}. So we take specific values of energy to the Hamiltonian $\mathcal{H}$ and make the diagram in the phase space of the dynamical system.

The figure \ref{fig1} shows the phase space of the solution \eqref{sol1} for the values $\mathcal{H}=\{\mathcal{H}_c-10^{-1},\mathcal{H}_c,\mathcal{H}_c+3\times 10^{-1}\}$, with $\mathcal{H}_c=0.25$. The values of the critical radius and critical velocity are obtained by the equation \eqref{critical}, resulting in $r_c=\sqrt{2r_0}=r_H$ e $v_c=1$, with $r_0=1$. We note here that the dynamic system \eqref{dotr}-\eqref{dotv} does not have really a critical point in the horizon, it is clear from the figure \ref{fig1}, where we have no intersection between the curves. The figure \ref{fig1} represent then the phase space of the dynamic system of the first solution, showing the fluid flow near the black hole event horizon. We observed from the figure that the fluid motion begins as purely subsonic accretion ($v>-v_c\equiv -1$), where in red curves represent the energy $\mathcal{H}_c-10^{-1}$, the blue curves the energy $\mathcal{H}_c$, and the green curves the energy $\mathcal{H}_c+3\times 10^{-1}$. This accretion indicates that how much further away from the horizon is the fluid, more slower the radial velocity of accretion. We see that the top of the diagram represents the ejection movement of the fluid. We also note that the closer the horizon is the fluid, faster it is ejects.           

\begin{figure}[h]
\centering
\begin{tabular}{rl}
\end{tabular}
\caption{\scriptsize{Representation of the phase space of the solution \eqref{sol1} for $\omega=1$.} }
\label{fig1}
\end{figure}

The behavior is exactly the same for all solutions reviewed here. It is easy to see that the Hamiltonian \eqref{H2} admits global solution to the radial velocity $v$, with value $v_{\infty}=0$ ($r\rightarrow \infty$). 

\subsection{Ultra-relativistic fluid} \label{subsec2}

The ultra-relativistic fluid can be classified by $p=(1/2)e$, then $\omega=1/2$.  The Hamiltonian \eqref{Hf}, for the solutions \eqref{sol1}-\eqref{sol4-1} is given by
\begin{eqnarray}
&&\mathcal{H}_1(r,v)=\frac{1}{r^2v}\sqrt{\frac{1}{(1-v^2)}\left(1-\frac{2r_0}{r^2}\right)}\,,\,\mathcal{H}_2(r,v)=\frac{1}{r^2v}\sqrt{\frac{\sqrt{r^4-4e^{r_0}}}{(1-v^2)r^2}}\label{H3}\,,\\
&&\mathcal{H}_3(r,v)=\frac{1}{r^2v}\sqrt{\frac{(r^6-2r_0)^{1/3}}{(1-v^2)r^2}}\,,\,\mathcal{H}_4(r,v)=\frac{1}{r^2v}\sqrt{\frac{1}{(1-v^2)(e^{r_0/2}r-1)^2}}\label{H4}\,.
\end{eqnarray}

The critical speed for the four solutions is always given by $v_c=\sqrt{\omega}=0.707107$. The horizons are given by \eqref{horizon}. The critical radius is determined by the equation \eqref{critical}, which for the solutions \eqref{sol1}-\eqref{sol4-1} we have
\begin{eqnarray}
r_{c(1)}=\sqrt{3r_0}\,,\,r_{c(2)}=\left(6e^{r_0}\right)^{1/4}\,,\,r_{c(3)}=\left(3r_0\right)^{1/6}\,,\,r_{c(4)}=\frac{2}{3}e^{-r_0/2}\,.
\end{eqnarray}
Critical values for the Hamiltonian are obtained for $\mathcal{H}(r_{c(i)},v_c)=\mathcal{H}_{c(i)}$ in \eqref{H3} and \eqref{H4}, with $i=1,...,4$. We represent the contour, or phase space, to the solutions taking fixed the Hamiltonian. For the first and fourth solutions we take $\mathcal{H}^2_{1}(r,v)=\{\mathcal{H}_{c(1)}^2-10^{-2},\mathcal{H}_{c(1)}^2,\mathcal{H}_{c(1)}^2+3\times 10^{-2}\}$ and $\mathcal{H}^2_{4}(r,v)=\{\mathcal{H}_{c(4)}^2-1000,\mathcal{H}_{c(4)}^2,\mathcal{H}_{c(4)}^2+1200\}$. Already the contours for the solution \eqref{sol2} are obtained for $\mathcal{H}^4_{2}(r,v)=\{\mathcal{H}_{c(2)}^4-10^{-3},\mathcal{H}_{c(2)}^4,\mathcal{H}_{c(2)}^4+3\times 10^{-3}\}$, and for solution \eqref{sol3} by the expression $\mathcal{H}^6_{3}(r,v)=\{\mathcal{H}_{c(3)}^6-10^{-1},\mathcal{H}_{c(3)}^6,\mathcal{H}_{c(3)}^6+2\}$.

We represent the phase space of the solutions in figure \ref{fig2}. In the diagram at the top of the left side is the phase space of the first solution \eqref{sol1}, we have the following types of movement of the fluid: a) red curves. Those who are before the critical radius, $r_{c(1)}=1.73205$, it begins with a supersonic ejection movement ($1>v>v_c$), going to subsonic ejection ($v_c>v>0$), coming close to the horizon with zero speed, then passing to the movement of a subsonic accretion ($0<v<-v_c$), ending in a supersonic accretion ($-v_c>v>-1$).   The curve in the lower part of the diagram begins as an supersonic accretion, passing the accretion for subsonic, ending with zero velocity away from the event horizon. The curve at the top of the diagram begins far from the horizon like a subsonic ejection, reaching supersonic ejection away from the horizon to near the speed of light. b) the blue curves. We have three possibilities here. The first is the curve of the region smaller than the critical radius. This movement begins as a supersonic ejection, it passes through a critical point of bifurcation unstable, becoming subsonic ejection, which reaches a zero speed on the horizon, becoming subsonic accretion, passing again to an unstable bifurcation, finishing as accretion supersonic. The second is that the bottom of the diagram, which begins as a supersonic accretion, through a bifurcation ending as a subsonic accretion. The third is the curve of the upper diagram, which begins as a subsonic ejection, through a bifurcation and ending as a supersonic ejection. c) the green curves. We have three types of movements here. At the bottom we have a purely super sonic accretion movement. At the top we have a purely supersonic ejection movement. Far from the horizon, near zero speed, fluid movement begins as subsonic ejection, increasing the speed to maximum, then decreasing the radial velocity to zero by passing the motion to subsonic accretion, and the farther away the horizon speed tends to zero.    
\begin{figure}[h]
\centering
\begin{tabular}{rl}
\end{tabular}
\caption{\scriptsize{Representation of the phase space for solutions \eqref{sol1} (top left), \eqref{sol2} (top right), \eqref{sol3} (lower left) and \eqref{sol4-1} (bottom right). The sound speed value is given by $a^2=\omega=1/2$.} }
\label{fig2}
\end{figure}

The movement of fluid to the solutions \eqref{sol2} and \eqref{sol3}  is identical to the first solution. The only differences are the values of the horizon and the critical radius, which are $\{r_{c(2)}=2.00961,r_{h(2)}=1.81589\}$ and $\{r_{c(3)}=1.20094,r_{h(3)}=1.12246\}$. The movement of the fluid for solution \eqref{sol4-1} is quite different from other solutions. The phase space for this solution is shown in figure \ref{fig2} at the bottom right. The critical radius and the horizon are located in $\{r_{c(4)}=0.404354,r_{h(4)}=0.606531\}$. We note that the value of the horizon this time is greater than the critical radius, making it clear in the diagram. We can see that we have the following movements for the fluid: a) the red curves, the fluid begins with supersonic accretion, passing for a subsonic accretion. Then, the top of the diagram begins as subsonic ejection,  and ending supersonic ejection. b) the blue curves, whose movement is identical to the red curves. c) the green curves. The curves in the region $r>r_h$ have the same movement of fluid to the curves in red. But the fluid exhibits a cyclic motion within the horizon, where there is a stable critical point. This movement can not be observed by an observer outside the event horizon. This is the first time it appears such a phenomenon.

\subsection{Radiation}\label{subsec3}

An important model for a fluid is when its total energy density is equivalent to one third pressure, then classify as radiation, and $\omega=1/3$. The Hamiltonians, arising from \eqref{Hf}, for this case are
\begin{eqnarray}
&&\mathcal{H}_1(r,v)=\left[\frac{(r^2-2r_0)}{r^4v(1-v^2)}\right]^{2/3}\,,\,\mathcal{H}_2(r,v)=\left[\frac{\sqrt{r^4-4e^{r_0}}}{r^4v(1-v^2)}\right]^{2/3}\label{H5}\,,\\
&&\mathcal{H}_3(r,v)=\left[\frac{(r^6-2r_0)^{1/3}}{r^4v(1-v^2)}\right]^{2/3}\,,\,\mathcal{H}_4(r,v)=\left[\left(e^{r_0/2}r-1\right)^2(1-v^2)r^2v\right]^{-2/3}\label{H6}\,.
\end{eqnarray}

The horizons are given by \eqref{horizon}. The critical rays are determined by \eqref{critical}
\begin{eqnarray}
r_{c(1)}=2\sqrt{r_0}\,,\,r_{c(2)}=\left(8e^{r_0}\right)^{1/4}\,,\,r_{c(3)}\left(2\sqrt{r_0}\right)^{1/3}=\,,\,r_{c(4)}=e^{-r_0/2}/2\,.
\end{eqnarray}
The critical speed of sound is given by $v_c=\sqrt{\omega}=0.57735$. 
\par 
The movements of the fluid for the solutions \eqref{sol2} and \eqref{sol4-1} are similar to the previous section, obtained the contour of the equations $\mathcal{H}_2^3(r,v)=\left\{\mathcal{H}_{c(2)}^3-10^{-2},\mathcal{H}_{c(2)}^3,\mathcal{H}_{c(2)}^3+3\times 10^{-2}\right\}$ and $\mathcal{H}_4^3(r,v)=\left\{\mathcal{H}_{c(4)}^3-10^{4},\mathcal{H}_{c(4)}^3,\mathcal{H}_{c(4)}^3+11\times 10^{3}\right\}$, with $\mathcal{H}_{c(2)}^3=0.155199,\mathcal{H}_{c(4)}^3=12768.3$. The figure \ref{fig3} shows the phase space of these solutions on the right side in the top and lower. As the movements are similar to the previous section, we will not comment on them here. The main difference is the critical radius for each movement.
\begin{figure}[h]
\centering
\begin{tabular}{rl}
\end{tabular}
\caption{\scriptsize{Representation of the phase space for solutions \eqref{sol1} (top left), \eqref{sol2} (top right), \eqref{sol3} (lower left) and \eqref{sol4-1} (bottom right). The sound speed value is given by $a^2=\omega=1/3$.} }
\label{fig3}
\end{figure}

In the case of the left side of the phase diagram at the top of the figure \ref{fig3}, concerning the equation $\mathcal{H}_1^3(r,v)=\left\{\mathcal{H}_{c(1)}^3-10^{-1},\mathcal{H}_{c(1)}^3,\mathcal{H}_{c(1)}^3+3\times 10^{-1}\right\}$ where $\mathcal{H}_{c(1)}^3=0.105469$, the movements are divided as follows: a) curves on red. The movement begins as supersonic ejection, passing to subsonic ejection, reducing the ejection speed to zero, then going to subsonic accretion, and ending as supersonic accretion. Interestingly, there is a movement inside the event horizon inversely to the outside, starting as supersonic accretion and ending as ejection (inside) supersonic. b) curves in blue, we have the beginning as supersonic accretion, through an unstable critical point, becoming accretion subsonic. At the top is similar. The movement begins as subsonic ejection, goes through an unstable critical point, ends as ejection supersonic. A novelty is the movement that appears inside the event horizon. This movement started as supersonic accretion, it passes subsonic accretion, speed reduces to zero, becoming subsonic ejection, and ending supersonic ejection. c) curves on green. There is an accretion of movement and other purely supersonic ejection. Two other movements, an accretion and other ejection purely subsonic. Similarly to other curves, there is a movement within the horizon that begins as supersonic accretion and ends as supersonic ejection. In the case of movements to the equation $\mathcal{H}_3^9(r,v)=\left\{\mathcal{H}_{c(3)}^9-1,\mathcal{H}_{c(3)}^9,\mathcal{H}_{c(3)}^9+2\right\}$, where $\mathcal{H}_{c(3)}^9=4.80542$, we have a lot of similarity with the previous case, in which the main difference is the existence of motion where the radial coordinate is greater than the critical radius, for curves in red, as shown in figure \ref{fig3}.

\subsection{Sub-relativistic fluid}\label{subsec4}

Now finally, we discuss the case where the pressure is equal to $p=\rho/4$, then $a^2=\omega=1/4$. This value of the speed of sound characterizes fluid as sub-relativistic. In this case, the Hamiltonian for the solutions \eqref{sol1}-\eqref{sol4-1} are given by 
\begin{eqnarray}
&&\mathcal{H}_1(r,v)=\frac{1}{r^2\sqrt{v}}\left[\frac{(r^2-2r_0)}{r^2(1-v^2)}\right]^{3/4}\,,\,\mathcal{H}_2(r,v)=\frac{1}{r^2\sqrt{v}}\left[\frac{\sqrt{r^4-4e^{r_0}}}{r^2v(1-v^2)}\right]^{3/4}\label{H7}\,,\\
&&\mathcal{H}_3(r,v)=\frac{1}{r^2\sqrt{v}}\left[\frac{(r^6-2r_0)^{1/3}}{r^2v(1-v^2)}\right]^{3/4}\,,\,\mathcal{H}_4(r,v)=\frac{1}{r^2\sqrt{v}}\left[\left(e^{r_0/2}r-1\right)^2(1-v^2)\right]^{-3/4}\label{H8}\,.
\end{eqnarray}

The horizons are given by \eqref{horizon}. The critics rays are determined by \eqref{critical}
\begin{eqnarray}
r_{c(1)}=\sqrt{5r_0}\,,\,r_{c(2)}=\left(10e^{r_0}\right)^{1/4}\,,\,r_{c(3)}=\left(5\sqrt{r_0}\right)^{1/6}\,,\,r_{c(4)}=\frac{2}{5}e^{-r_0/2}\,.
\end{eqnarray}
The critical speed of sound is given by $v_c=\sqrt{\omega}=0.5$. The figure \ref{fig4} shows the phase space of Hamiltonians \eqref{H7} and \eqref{H8}. The contours were obtained for the equations $\mathcal{H}_1^4(r,v)=\left\{\mathcal{H}_{c(1)}^4-10^{-2},\mathcal{H}_{c(1)}^4,\mathcal{H}_{c(1)}^4+3\times 10^{-2}\right\}$ where $\mathcal{H}_{c(1)}^4=0.08192$; $\mathcal{H}_2^8(r,v)=\left\{\mathcal{H}_{c(2)}^8-10^{-3},\mathcal{H}_{c(2)}^8,\mathcal{H}_{c(2)}^8+3\times 10^{-2}\right\}$ where $\mathcal{H}_{c(2)}^8=0.0262795$;  $\mathcal{H}_3^4(r,v)=\left\{\mathcal{H}_{c(3)}^4-0.8,\mathcal{H}_{c(3)}^4,\mathcal{H}_{c(3)}^4+2\right\}$ where $\mathcal{H}_{c(3)}^4=1.94557$; and $\mathcal{H}_4^4(r,v)=\left\{\mathcal{H}_{c(4)}^4-5\times 10^{4},\mathcal{H}_{c(4)}^4,\mathcal{H}_{c(4)}^4+51\times 10^{3}\right\}$ where $\mathcal{H}_{c(4)}^4=58656.7$. We can clearly see the figure \ref{fig4}, that the movements of the fluid are similar to those described in subsection \ref{subsec2}, so we will not specify them here. 

\begin{figure}[h]
\centering
\begin{tabular}{rl}
\end{tabular}
\caption{\scriptsize{Representation of the phase space for solutions \eqref{sol1} (top left), \eqref{sol2} (top right), \eqref{sol3} (lower left) and \eqref{sol4-1} (bottom right). The sound speed value is given by $a^2=\omega=1/4$.} }
\label{fig4}
\end{figure}

\section{Conclusion}\label{sec3}

The spherical accretion of matter by charged black holes on $ f (T) $ Gravity is studied and the conclusion is that it is possible, resulting in several movements for the fluid.
\par 
Considering a perfect isentropic and isothermal fluid, we analyzed the cases of ultra-stiff, ultra-relativistic, radiation and sub-relativistic fluids. We consider the contour for each specific value of the Hamiltonian as a constant movement, and represent the phase space for the four analyzed solutions. The conclusion follows that it is perfectly possible accretion and ejection of matter spherical type-fluid by the black holes studied. The movement of the fluid more deserves emphasis here is the cyclic appearing within the horizon for a given energy of the fluid.
\par 
Our perspective is that these solutions are stable for both a thermodynamic system (thermodynamic stability) and for small perturbations in the geometry (geometric stability). This should be the next test for these solutions in future work.
 

\vspace{1cm}

{\bf Acknowledgement}: MER thanks UFPA, Edital 04/2014 PROPESP, and CNPq,
Edital MCTI/CNPQ/Universal 14/2014,  for partial financial support.


%
\end{document}